\pdfoutput=1
\documentclass[a4paper,reprint,twocolumn,notitlepage,aip,nofootinbib]{revtex4-2}

\usepackage[left=1.5cm,right=1.5cm,top=1cm,bottom=1.5cm,includeheadfoot]{geometry}

\usepackage{tgtermes} 
\usepackage{tgheros} 
\usepackage[T1]{fontenc}

\usepackage{amssymb}
\usepackage{amsmath}
\usepackage{amsthm}

\usepackage{graphicx}
\usepackage{ushort}
\usepackage{bm} 
\PassOptionsToPackage{hyphens}{url} 
\usepackage[breaklinks=true]{hyperref}
\usepackage[dvipsnames]{xcolor}
\usepackage{soul}

\let\oldAA\AA
\newcommand{\Angstrom}{\text{\normalfont\oldAA}}

\newcommand{\R}{\mathbb{R}}

\newcommand{\xx}{\mathbf{x}}
\newcommand{\rr}{\mathbf{r}}
\newcommand{\jj}{\mathbf{j}}

\newcommand{\FF}{\mathbf{F}}

\renewcommand{\AA}{\mathbf{A}}

\newcommand{\alphafx}{\bm{\alpha}_\mathrm{fx}}
\newcommand{\alphafc}{\bm{\alpha}_\mathrm{fc}}

\newcommand{\vx}{v_\mathrm{x}}
\newcommand{\vfx}{v_\mathrm{fx}}
\newcommand{\vH}{v_\mathrm{H}}

\newcommand{\vHxc}{v_\mathrm{Hxc}}
\newcommand{\vfHx}{v_{\rm fHx}}
\newcommand{\vfc}{v_{\rm fc}}

\newcommand{\FH}{\FF_\mathrm{H}}
\newcommand{\Fx}{\FF_\mathrm{x}}
\newcommand{\Fc}{\FF_\mathrm{c}}
\newcommand{\FHxc}{\FF_\mathrm{Hxc}}
\newcommand{\Ex}{E_\mathrm{x}}
\newcommand{\ff}{\boldsymbol{f}} 
\newcommand{\fx}{\ff_\mathrm{x}}
\newcommand{\fc}{\ff_\mathrm{c}}
\newcommand{\fHxc}{\ff_\mathrm{Hxc}}

\renewcommand{\d}{\,\mathrm{d}} 

\let\Im\relax
\DeclareMathOperator{\Im}{Im}

\def\brakett#1#2#3{\langle #1 | \,#2\, | #3\rangle}

\begin{document}

\title{Exchange energies with forces in density-functional theory}

\author{Nicolas Tancogne-Dejean}
\email[Electronic address:\;]{nicolas.tancogne-dejean@mpsd.mpg.de}
\affiliation{Max Planck Institute for the Structure and Dynamics of Matter and Center for Free-Electron Laser Science \& Department of Physics, Hamburg, Germany}

\author{Markus Penz}
\affiliation{Department of Computer Science, Oslo Metropolitan University, Norway}
\affiliation{Basic Research Community for Physics, Innsbruck, Austria}

\author{Andre Laestadius}
\affiliation{Department of Computer Science, Oslo Metropolitan University, Norway}
\affiliation{Hylleraas Centre for Quantum Molecular Sciences, Department of Chemistry, University of Oslo, Norway}

\author{Mih\'aly A. Csirik}
\affiliation{Department of Computer Science, Oslo Metropolitan University, Norway}
\affiliation{Hylleraas Centre for Quantum Molecular Sciences, Department of Chemistry, University of Oslo, Norway}

\author{Michael Ruggenthaler}
\affiliation{Max Planck Institute for the Structure and Dynamics of Matter and Center for Free-Electron Laser Science \& Department of Physics, Hamburg, Germany}
\affiliation{The Hamburg Center for Ultrafast Imaging, Hamburg, Germany}

\author{Angel Rubio}
\affiliation{Max Planck Institute for the Structure and Dynamics of Matter and Center for Free-Electron Laser Science \& Department of Physics, Hamburg, Germany}
\affiliation{Center for Computational Quantum Physics, Flatiron Institute, New York, USA}
\affiliation{The Hamburg Center for Ultrafast Imaging, Hamburg, Germany}

\begin{abstract}
We propose exchanging the energy functionals in ground-state DFT with physically equivalent exact force expressions as a new promising route towards approximations to the exchange-correlation potential and energy. In analogy to the usual energy-based procedure, we split the force difference between the interacting and auxiliary Kohn--Sham system into a Hartree, an exchange, and a correlation force. The corresponding scalar potential is obtained by solving a Poisson equation, while an additional transverse part of the force yields a vector potential. These vector potentials obey an exact constraint between the exchange and correlation contribution and can further be related to the atomic-shell structure.
Numerically, the force-based local-exchange potential and the corresponding exchange energy compare well with the numerically more involved optimized effective-potential method. Overall, the force-based method has several benefits when compared to the usual energy-based approach and opens a route towards numerically inexpensive non-local and (in the time-dependent case) non-adiabatic approximations.
\end{abstract}

\maketitle

\section{Introduction}

It is with great pleasure that we provide this contribution to the special issue of the \emph{Journal of Chemical Physics}  honoring John Perdew and his work in quantum chemistry.
John Perdew has made groundbreaking advancements in developing exchange-correlation functionals within density-functional theory (DFT), which are crucial for the accurate description of the interactions between electrons.
DFT~\cite{dreizler2012}, with its many variants~\cite{engel2011,ullrich2011,marques2012fundamentals,ruggenthaler2015b,vignale1987density,vignale2004mapping,ruggenthaler2015,ruggenthaler2014}, is nowadays the workhorse of first-principle simulations in quantum chemistry, solid state physics and materials science, and John Perdew greatly contributed to this success story.
His work has focused on improving the accuracy and efficiency of density-functional calculations by proposing more precise and robust functionals, allowing researchers to study a wide range of chemical and physical properties of materials. Specifically, he has explored fundamental (exact) constraints that an exchange-correlation functional must satisfy to accurately describe the electronic interactions in a system. Enforcing such exact constraints can greatly improve the reliability of approximate functionals~\cite{Kaplan2023}. This work follows this general idea by taking up previous suggestions on how to rephrase the exchange-correlation potential in terms of forces. We give known and novel exact force-based constraints and show how to translate these ideas into an efficient numerical scheme.

Most DFT simulations are performed using the Kohn--Sham (KS) scheme~\cite{KS1965}, where the density of the interacting system is predicted by solving an auxiliary non-interacting system. It is precisely the mentioned exchange-correlation potential that relates the interacting and the non-interacting system through the underlying density-potential mapping $v(\rr) \leftrightarrow \rho(\rr)$. For a recent review on this mapping in the context of DFT, we point to~\citet{Penz-et-al-HKreview-partI}.
It is common practice to derive approximations for the (in general unknown) exchange-correlation potential by re-expressing the universal density functional as a sum of non-interacting kinetic, Hartree and exchange-energy functionals, as well as the unknown correlation-energy functional~\cite{burke2012perspective} and then to assume functional differentiability~\cite{blanchard2015} with respect to the density. While for approximate functionals that are given explicitly in terms of the density, potentials can be determined this way by direct differentiation, for implicit functionals this is no longer possible in general~\cite{van2003density}. To make matters worse, it has been shown that the universal density functional of DFT is not functionally differentiable with respect to the usual function spaces~\cite{lammert2007differentiability}. While a generalized definition of functional differentiability (subdifferentiability) is enough to establish the mapping from $v$-representable densities to potentials~\cite{Lieb1983Density}, many of the commonly employed rules of differential calculus, such as linearity or the chain rule, might no longer hold in the same way~\cite{clason-valkonen-book}. This fact therefore questions this common way to infer exchange-correlation potentials from exchange-correlation energy functionals. We note that the theoretical setting of an exact regularization procedure is available that renders the involved functionals differentiable~\cite{kvaal2014differentiable,KSpaper2018,laestadius2019kohn} and that this surprisingly links to the Zhao--Morrison--Parr method for mapping $\rho(\rr)\mapsto v(\rr)$ \cite{penz2023-ZMP}.
Importantly, in this work we highlight that also the exchange-only energy is non-differentiable with respect to densities, thus allowing \emph{local}-exchange potentials only in the form of generalized constructions such as the optimized effective potential (OEP), or, alternatively, leading to an additional vector potential for exchange effects. This vector potential naturally appears in a force-based approach and acts semi-locally on the wave function. This is in contrast to the exchange term in Hartree--Fock that acts fully non-locally on the wave functions.

From a physical point of view, one can always exchange the description of a many-particle quantum system in its ground state in terms of energies by a description based on forces. Both views have been viable routes towards getting the desired potential. Indeed, the exact exchange-correlation potential of DFT can be expressed directly in terms of the difference in force densities between the interacting and the auxiliary non-interacting system~\cite{holas-march1995,holas-march-rubio-2005,ruggenthaler2009local,tchenkoue2019force}, thus bypassing functional differentiation and related issues.
A method for deriving DFT potentials from the electric field due to the Fermi-Coulomb-hole charge distribution was pioneered by Harbola and coworkers \cite{Harbola-Sahni,harbola1991local,sahni1992atomic,slamet1994force}. However, this approach misses the kinetic-correlation contribution and thus does not retrieve the full exchange-correlation potential \cite{nagy1990interpretation}. This issue was also noted by \citet{holas-march1995}, who first used a force-based approach to give an expression for the exchange-correlation potential of DFT in the form of a (path-independent) line integral. Building upon this important work, \citet{sahni1997physical} was able to extend the method of Harbola.

In this work, we show that a force-based approach is not only conceptually very appealing but also practically relevant.
In doing so, we stick to a fully spin-resolved, collinear formulation.
Specifically, we show that besides the usual Hartree potential, we can also derive the simple explicit form of the local-exchange potential previously suggested by \citet{Harbola-Sahni}. This potential we show to be directly linked to the exchange force density and it enters a generalized exchange virial relation.
A different form of a generalized exchange virial relation is actually discussed in another paper of this special edition.~\cite{xVR-special-edition-paper}
We further find a relation between the exchange and the correlation force densities that takes the form of a novel exact constraint. As we demonstrate, the formulation of the force-based local-exchange potential is consistent with current-density-functional theory (CDFT) \cite{vignale1987density,Penz-et-al-HKreview-partII} and we discuss its connection to the time-dependent case. In the context of ground-state DFT we then show that the explicit force-based local-exchange potential performs similarly to the numerically much more involved optimized effective potential (OEP) approach in exchange approximation. We show that the difference between OEP and force-based local-exchange potential can be connected to the above mentioned exact constraint that exchange and correlation force densities need to fulfill. We finally comment on practical ways on how to treat the remaining correlation force densities. In this we highlight how the force-based approach provides a route towards numerically inexpensive non-local (in how it depends on the density) and non-adiabatic functionals that also act semi-locally on the wave function if they contain a vector-potential contribution.

\section{Force-based Kohn--Sham setting}

To start with, we consider the $N$-particle Hamiltonian (in Hartree atomic units $e = \hbar = m_e = (4\pi\epsilon_0)^{-1} = 1$), first in a time-dependent setting while we later switch to ground states.
\begin{align}\label{eq:hamiltonian}
\hat{H} =\underbrace{-\frac{1}{2}\sum_{k=1}^N \nabla^2_k}_{\textstyle \hat{T}} + \underbrace{\sum_{k=1}^N v(\rr_k\sigma,t)}_{\textstyle \hat{V}[v]}  + \underbrace{\sum_{k>l} \frac{1}{|\rr_k-\rr_l|}}_{\textstyle \hat{W}}
\end{align}
Here, $v(\rr\sigma,t)$ is the external, spin-resolved one-particle potential at time $t$. 
Note that while the external potential can act separately on the spin components, we here do not take an external magnetic field nor spin-orbit coupling into account. We at the end comment how to extend the present force-based formalism to these cases as well.
For anti-symmetric wave functions $\Psi(\xx_1,...,\xx_N,t)$, where $\xx_k =(\rr_k \sigma_k)$, we define the spin-resolved $p^{\rm th}$-order reduced density matrix
\begin{equation}
\begin{aligned}
&\rho^{(p)}(\xx_1, \ldots, \xx_p,\xx_1', \ldots, \xx_p',t)\\
&= \frac{N!}{p!(N-p)!} \sum_{\substack{\sigma_{p+1} \ldots \sigma_N}}\int \Psi(\xx_1,\ldots,\xx_p,\xx_{p+1},\ldots,\xx_N,t)\\
& \quad \times\Psi^{*}(\xx'_1,\ldots,\xx_p',\xx_{p+1},\ldots,\xx_N,t) \d\rr_{p+1}\ldots\d\rr_N.
\label{p-RDM}
\end{aligned}
\end{equation}
We can then use these reduced density matrices and the spin-resolved density $\rho(\xx,t) = \rho(\rr\sigma,t) = \rho^{(1)}(\rr\sigma,\rr\sigma,t)$ to express the (paramagnetic and spin-resolved) current density 
\begin{align}
\jj(\rr\sigma,t) = \Im \left(\left.\nabla \rho^{(1)}(\rr\sigma,\rr'\sigma,t) \right|_{\rr'=\rr}  \right)
\end{align} 
and its equation of motion~\cite{stefanucci2013, tchenkoue2019force} (also called ``local force-balance equation'')
\begin{align}\label{eq:force-balance}
\partial_t \jj(\rr\sigma,t) = -\rho(\rr\sigma,t)\nabla v(\rr\sigma,t) + \FF_T(\rr\sigma,t) + \FF_W(\rr\sigma,t). 
\end{align}
This expression introduces the exact interaction-stress and momentum-stress force densities, respectively,
\begin{equation}\label{eq:interaction-force}
\begin{aligned}
\FF_W(\rr\sigma,t)=&- 2\sum_{\sigma'}\!\!\int\! (\nabla |\rr'-\rr|^{-1}) \\
&\times \rho^{(2)}(\rr\sigma,\rr'\sigma',\rr\sigma,\rr'\sigma',t) \d \rr',\\
\end{aligned}
\end{equation}
\begin{equation}\label{eq:momentum-stress-force}
\FF_T(\rr\sigma,t) = \left.\frac{1}{4} (\nabla - \nabla')(\nabla^2 - \nabla'^2)\rho^{(1)}(\rr\sigma,\rr'\sigma,t)\right|_{\rr'=\rr}. 
\end{equation}
Here, $(\nabla |\rr'-\rr|^{-1})$ indicates that the gradient only acts on the Coulomb interaction term.
Those force terms can be linked directly to the quantum stress tensor \cite{tokatly2005-1} that includes information about the atomic shell structure \cite{TaoVignaleTokatly2008}.
Equation~\eqref{eq:force-balance} has been the primary starting point for inquiries in time-dependent DFT (TDDFT). Among other things, it was used to provide a mapping from densities to potentials~\cite{van1999mapping}, to analyze features of the time-dependent exchange-correlation potential~\cite{Luo2014}, to get exact constraints as well as formulations for non-adiabatic approximate functionals~\cite{fuks2018exploring,Lacombe2019}, and to reformulate KS-TDDFT in terms of the second time derivative of the density~\cite{Tarantino2021}. While here we focus on the ground-state problem, some consequences for the time-dependent case will be discussed further in Sec.~\ref{sec:other}.

In the following, we indicate the terms coming from the solution $\Psi$ of the fully interacting problem as $\FF_W[\Psi]$ and $\FF_T[\Psi]$. The auxiliary, non-interacting KS problem is controlled by the Hamiltonian $\hat H_s = \hat T + \hat V[v_s]$, including a different external potential $v_s(\rr\sigma,t)$, and has a Slater-determinant solution $\Phi$. Analogous to Eq.~\eqref{eq:force-balance} we then have for the auxiliary system 
\begin{align}\label{eq:force-balance-s}
\partial_t \jj_s(\rr\sigma,t) = -\rho_s(\rr\sigma,t)\nabla v_s(\rr\sigma,t) + \FF_T[\Phi](\rr\sigma,t),
\end{align}
with a different current density $\jj_s$.

We now assume that all potentials are time-independent, that we are in the ground state for both systems, and further that they both generate the same ground-state density, i.e., $\rho(\rr\sigma) = \rho_s(\rr\sigma)$. In the ground state it also holds $\partial_t \jj(\rr\sigma) = \partial_t \jj_s(\rr\sigma) = 0$ and we find with the definition of the Hartree exchange-correlation (Hxc) potential $\vHxc(\rr\sigma) = v_s(\rr\sigma) - v(\rr\sigma)$ that
\begin{equation}\label{eq-Hxc-potential}
\rho\nabla \vHxc = -\FHxc[\Phi, \Psi] = \FF_T[\Phi] - \FF_T[\Psi] - \FF_W[\Psi],
\end{equation}
which defines $\FHxc$ for each spin channel.
By virtue of the Hohenberg--Kohn theorem \cite{hohenberg1964,Penz-et-al-HKreview-partI} and assuming non-degeneracy of the ground states for simplicity, the Slater determinant $\Phi$ as well as the the interacting wave function $\Psi$ are given solely and uniquely in terms of the density, which makes all the force densities determined by the density only. 
Equation~\eqref{eq-Hxc-potential} implies that
\begin{align}\label{eq:PurelyLongitundinal}
\nabla \vHxc(\rr\sigma) = -\frac{\FHxc[\Phi, \Psi](\rr\sigma)}{\rho(\rr\sigma)} = -\fHxc(\rr\sigma)
\end{align}
is a purely longitudinal (conservative) vector field. Since the Hartree contribution is longitudinal as well, so is the remaining exchange-correlation part. But if we decide to split the exchange-correlation part into its exchange and correlation contributions, as it is typically done for the energy, we do not have any such knowledge about these individual contributions any more. So the exchange and correlation vector fields can and will contain a non-zero transverse component.

Now, we can recast Eq.~\eqref{eq:PurelyLongitundinal} into a Poisson equation $\nabla^2 \vHxc = -\nabla \cdot \fHxc$ by applying the divergence and solve for for $\vHxc$ using the corresponding Green's function for the spatial domain $\R^3$,
\begin{align}\label{eq-Helmholtzexpression}
\vHxc(\rr\sigma) = \int \frac{\nabla' \cdot\fHxc(\rr'\sigma) }{4 \pi |\rr-\rr'|} \d\rr'.
\end{align}
Equation~\eqref{eq-Helmholtzexpression} represents the direct link between Hxc force density and the corresponding potential. Unlike the link between the energy and the potential, no functional differentiability is involved here.

Next, we split up the Hxc force density in analogy to the partition of the energy usual in DFT as
\begin{align}\label{eq:Hxcforces}
\FHxc[\Phi,\Psi] = \;& \FF_W[\Phi] \\
& + \underbrace{\FF_T[\Psi] -\FF_T[\Phi]  + \FF_W[\Psi] - \FF_W[\Phi]}_{\textstyle \Fc[\Phi,\Psi]},\nonumber
\end{align}
where $\FF_W[\Phi]$ is the Hartree-exchange (Hx) force density and $\Fc[\Phi,\Psi]$ the correlation force density. If desirable, the correlation part can be split again into a kinetic-correlation contribution $\FF_T[\Psi] -\FF_T[\Phi]$ and an interaction-correlation contribution $\FF_W[\Psi] - \FF_W[\Phi]$. The partition of Eq.~\eqref{eq:Hxcforces} leads to the respective force-based potentials, $\vfHx$ and $\vfc$, that add up to the exact Hxc potential,
\begin{align}\label{eq:HelmholtzHx}
\vHxc(\rr\sigma) = \underbrace{\int \frac{\nabla' \cdot\ff_{\rm Hx}(\rr'\sigma) }{4 \pi |\rr-\rr'|} \d\rr'}_{\textstyle \vfHx(\rr\sigma)} + \underbrace{\int \frac{\nabla' \cdot\ff_{\rm c}(\rr'\sigma) }{4 \pi |\rr-\rr'|} \d\rr'}_{\textstyle \vfc(\rr\sigma)}.
\end{align}
Here, we have denoted $\ff_{\rm Hx} = \FF_W[\Phi]/\rho$ and $\ff_{\rm c} = \Fc[\Phi,\Psi]/\rho$. Since the Hx force density is given in terms of the KS wave function only, we know this part explicitly and we can in principle calculate the exact force-based Hx potential for a given KS wave function.

To make the resulting force-based Hx potential more explicit, we make use of the fact that $\Phi$ is a single, closed-shell Slater determinant with spin-space orbitals $\varphi_{k}(\rr\sigma)$. We can then express~\cite{parr-yang-book} 
\begin{equation}
\begin{aligned}
\rho_s^{(2)}&(\rr\sigma,\rr'\sigma',\rr\sigma,\rr'\sigma') \\
&= \frac{1}{2}\left(\rho(\rr\sigma) \rho(\rr'\sigma') - \delta_{\sigma\sigma'} |\rho_s^{(1)}(\rr\sigma,\rr'\sigma')|^2 \right),
\end{aligned}
\end{equation}
where $\rho_s^{(1)}(\rr\sigma,\rr'\sigma') = \sum_{k} \varphi_{k}(\rr \sigma)\varphi^{*}_{k}(\rr' \sigma')$. Therefore the Hx force density splits naturally into a Hartree and an exchange term,
\begin{equation}\label{eq-interaction-force-single-slater}
	\begin{aligned}
	\FF_W[\Phi] = &\FH[\Phi] + \Fx[\Phi] = - \rho(\rr\sigma)\nabla \sum_{\sigma'} \! \int \! \frac{\rho(\rr'\sigma')}{|\rr-\rr'|} \d\rr' \\ 
	&+ \int (\nabla |\rr-\rr'|^{-1}) |\rho_s^{(1)}(\rr\sigma,\rr'\sigma)|^2 \d\rr'.
	\end{aligned}
\end{equation}
Note that while the Hartree mean-field acts on both spin channels, the exchange force density only links to the same spin component.
If $\Phi$ would be the Slater determinant from a non-local Hartree--Fock calculation then these terms would be the corresponding Hartree and Fock exchange force densities, respectively.
From the Hartree force density $\FH(\rr\sigma) = -\rho(\rr\sigma)\nabla \vH(\rr)$ we read off the (spin-summed) Hartree potential 
\begin{align}\label{eq:vHartree}
v_{\rm H}(\rr) = \sum_{\sigma'} \! \int \! \frac{\rho(\rr'\sigma')}{|\rr-\rr'|} \d \rr'.
\end{align}
The potential from the exchange terms will be derived in Sec.~\ref{sec:discussion-vfx}. The exchange force density satisfies an exchange virial relation that gives the exchange energy (see App.~\ref{app:VirialExchange} for details)
\begin{equation}
     \Ex[\Phi] = \sum_{\sigma} \! \int \! \rr \cdot \Fx[\Phi](\rr\sigma) \d\rr  .
\end{equation}
The exchange force density can be interpreted as the force on a test particle in the electric field of the exchange hole, as detailed in \citet{Harbola-Sahni}. 
This relation provides an important link from forces, or approximations to them, back to the respective energies.

\section{Force-based exact constraints}

Let us now comment on some exact constraints for the force densities. 
If, for the sake of consistency, just a single particle with wave function $\varphi(\rr\sigma)$ is considered, then it directly follows $\FF_W[\varphi]=0$ (no self-force) and naturally $\Fc[\varphi,\varphi]=0$. Note that $\FF_W[\varphi]=0$ can also be deduced from Eq.~\eqref{eq-interaction-force-single-slater}. Also, in the one-particle case, because of $\Psi=\Phi=\varphi$ the kinetic-correlation and the interaction-correlation force densities must vanish independently.
We further remark that these self-interaction properties are directly related to the corresponding expressions for the energy, which serve as a basis for the construction of self-interaction corrections, as pioneered by \citet{PhysRevB.23.5048}. A similar scheme could thus be developed on the basis of forces.

The zero-force and zero-torque constraints~\cite{levy1985hellmann} in the force-based formulation for the ground state take the simple form
\begin{equation}\label{eq:zero-force-and-torque}
    \sum_{\sigma} \! \int \! \FHxc(\rr\sigma) \d\rr = 0,\quad
    \sum_{\sigma} \! \int \! \rr\times\FHxc(\rr\sigma) \d\rr = 0.
\end{equation}
They even hold for each spin channel independently with a Hamiltonian like Eq.~\eqref{eq:hamiltonian} that does not feature any non-collinear magnetism.
Since we have an explicit expression for the contribution $\FF_W[\Phi]$ to the full $\FHxc[\Phi,\Psi]$ available, we can tighten these constraints further. 
Equation~\eqref{eq-interaction-force-single-slater} yields an anti-symmetric integrand in $\rr\sigma,\rr'\sigma'$ in Eq.~\eqref{eq:zero-force-and-torque} that must be invariant under the exchange of $\rr\sigma\leftrightarrow\rr'\sigma'$. This means that Eq.~\eqref{eq:zero-force-and-torque} holds for $\FF_W[\Phi]$ independently and thus we also receive exact constraints for just the correlation force,
\begin{equation}
    \sum_{\sigma} \! \int \! \Fc(\rr\sigma) \d\rr = 0,\quad
    \sum_{\sigma} \! \int \! \rr\times\Fc(\rr\sigma) \d\rr = 0.
\end{equation}
This property can even be independently formulated for the kinetic-correlation and the interaction-correlation force densities, as shown in~\citet{fuks2018exploring}.

A further exact constraint that holds locally for the exchange and correlation vector fields is derived at the end of Section~\ref{sec:discussion-vfx}.

\section{Discussion of the force-based local-exchange potential}
\label{sec:discussion-vfx} 

Let us now consider how to make the above relations between force densities practical for DFT applications.
Using Eqs.~\eqref{eq:HelmholtzHx} and \eqref{eq-interaction-force-single-slater}, we can define the force-based local-exchange potential
\begin{align}
\label{eq:LocalExchangePotential-1}
\vfx(\rr\sigma) = -\int \frac{\nabla' \cdot \int (\nabla' |\rr''-\rr'|^{-1})\bar{\rho}_{\rm x}(\rr''|\rr'\sigma)\d\rr''}{4 \pi |\rr-\rr'|}\d\rr'
\end{align}
that together with the Hartree term gives $\vfHx = \vH + \vfx$.
Here, we used the usual definition of the exchange-hole density~\cite{parr-yang-book} (without a factor $\frac{1}{2}$ since it is spin-resolved)
\begin{align}
\bar{\rho}_{\rm x}(\rr'|\rr\sigma)= -\frac{|\rho_s^{(1)}(\rr\sigma,\rr'\sigma)|^2}{\rho(\rr\sigma)}.
\end{align}

The potential $\vfx$ is therefore the exchange potential that originates from only the \textit{longitudinal part} of the exchange vector field $\fx = \Fx[\Phi]/\rho$. We will come back to this point and its implications for density-functional approximations below.
To complete the picture, the missing correlation potential is given uniquely in terms of the (unknown) force-density difference $\FF_{\rm c}[\Phi,\Psi]$ from Eq.~\eqref{eq:Hxcforces} and a simple Coulomb integral (see Eq.~\eqref{eq:HelmholtzHx}).
For the force-based local-exchange potential given by Eq.~\eqref{eq:LocalExchangePotential-1} a numerically more convenient form in terms of the Slater-exchange potential plus correction terms can be derived (see App.~\ref{app:Convenient}).
It also obeys the usual coordinate scaling relations (see App.~\ref{app:VirialExchange}).

Based on the above explicit form of the local-exchange potential, we can highlight differences to the usual energy-based approach and point out potential advantages of the force-based approach. In the energy-based approach, the potential is found via a functional variation of the energy expression with respect to the density. In the exchange case one considers the functional derivative of
\begin{equation}
\begin{aligned}\label{eq:ExchangeEnergy}
\Ex[\rho] & = \brakett{\Phi[\rho]}{\hat{W}}{ \Phi[\rho]} - E_{\rm H}[\rho] \\
&= -\frac{1}{2} \sum_\sigma\!\int \frac{ |\rho_s^{(1)}(\rr\sigma,\rr'\sigma)|^2}{|\rr-\rr'|} \d\rr \d\rr',
\end{aligned}
\end{equation}
where $E_{\rm H}[\rho] = \frac{1}{2} \sum_\sigma\int \vH(\rr) \rho(\rr\sigma) \d\rr$ is the Hartree energy.
Now, $\Ex[\rho]$ is defined as a \emph{density} functional by invoking the usual mapping $\rho\mapsto\Phi$.
As was pointed out by \citet{van2003density}, for an implicit density functional the (generalization of the) functional derivative is not straightforward and does not exist in general. On the other hand, if the functional derivative would exist, then \emph{by construction} it obeys a virial relation of the form (see App.~\ref{app:VirialExchange})
\begin{align}\label{eq:virial-relation-standard}
 \Ex[\rho] = - \sum_\sigma \! \int\! \rho(\rr\sigma)\, \rr \cdot \nabla \frac{\delta \Ex[\rho]}{\delta \rho (\rr\sigma)} \d\rr.    
\end{align}
In practice, the derivative is determined by the OEP approach~\cite{SHARP_PR90_317,TALMAN_PRA14_36} that needs to assume Fr\'echet (total functional) differentiability to allow for the application of the functional chain rule~\cite{krieger1992systematic,KuemmelPerdew2003}.
Yet, OEP exchange potentials, in accordance with non-differentiability of the exchange-energy functional, in general do not obey Eq.~\eqref{eq:virial-relation-standard}. Sometimes this relation is additionally imposed, e.g., in \citet{PhysRevA.57.3425} (also compare Tab.~\ref{tab:table_virial}), however this will not restore differentiability. Consequently, the OEP procedure needs to be interpreted as a local-potential approximation, but not to an actually existing local-exchange potential defined by a functional derivative.
In the force-based approach, that avoids any reference to functional differentiability, a different virial relation is derived. We start by applying the Helmholtz decomposition~\cite{arfken-book}  to the exchange vector field. This yields a longitudinal (curl-free) and a transverse (divergence-free) vector-field component.
\begin{align}\label{eq:HelmholtzDecomposition}
    \fx(\rr\sigma) = \frac{\Fx[\Phi](\rr\sigma)}{\rho(\rr\sigma)} = - \nabla \vfx(\rr\sigma) + \nabla \times \alphafx(\rr\sigma)
\end{align}
With this we find the generalized exchange virial relation (see App.~\ref{app:VirialExchange} for details; \citet{harbola1991local} give the same relation, just without spin sum)
\begin{equation}\label{eq:VirialGeneral}
\begin{aligned}
    \Ex[\Phi] =& \sum_\sigma \! \int\! \rr \cdot \Fx[\Phi](\rr\sigma) \d\rr 
    \\
    =& -\sum_\sigma \! \int\!\rho(\rr\sigma) \, \rr \cdot \nabla \vfx(\rr\sigma) \d\rr \\
    &+ \sum_\sigma \! \int\! \rho(\rr\sigma) \, \rr \cdot (\nabla \times \boldsymbol{\alpha}_{\rm fx}(\rr\sigma)) \d\rr.
\end{aligned}
\end{equation}
The last term due to the curl does vanish for spherically-symmetric densities (see App.~\ref{app:VirialExchange}, where we also give an explicit formula for $\alphafx$, and Tab.~\ref{tab:table_virial}). Hence, $\vfx$ satisfies a virial relation of the form of Eq.~\eqref{eq:virial-relation-standard} for closed-shell spherically-symmetric systems, but in general we have the more involved Eq.~\eqref{eq:VirialGeneral} including a transverse component through the curl term.
This is due to the fact that the exchange vector field $\fx$ is \emph{not purely longitudinal}, and hence while the exchange energy is directly linked to the exchange force density, the longitudinal part of the exchange vector field alone cannot yield the full exchange energy in general.
Since OEP methods do not fulfill the virial relation of the form of Eq.~\eqref{eq:virial-relation-standard} even in the spherically-symmetric case (see Tab.~\ref{tab:table_virial}), this implies that the local-exchange potential from the force-balance approach is in general different from an exchange potential defined as a (generalized) exchange-energy derivative~\cite{van2003density}, like obtained by common OEP procedures.
This was already pointed out by \citet{wang1990exchange} when discussing the local-exchange potential of \citet{Harbola-Sahni}, which is equivalent to Eq.~\eqref{eq:LocalExchangePotential-1}. To show this, they derived the second-order gradient expansion of both, the gradient of the energy-based exchange potential and $\fx = \Fx[\Phi]/\rho$ from Eq.~\eqref{eq-interaction-force-single-slater}, and showed that the expressions do not match. But note that in order to make this a strict statement about the potentials, we need to assume that $\fx$ is a gradient field, i.e., $\alphafx=0$, which does not hold in general.

On the practical side, if one is only interested in finding a local potential that minimizes the exchange energy then the common OEP approaches will usually perform better than the force-based local-exchange potential (see Tab.~\ref{tab:table_HF} in App.~\ref{app:Numerical} for comparison to Hartree--Fock results). This is by design, since the exchange-only OEP procedure is precisely such that it seeks the \textit{local} potential $v_\mathrm{OEPx}$ that minimizes the energy with an uncorrelated state~\cite{Yang2002OEP}. Herein, the state is always a Slater determinant constructed from the orbitals of a one-particle Hamiltonian with the chosen potential $v_\mathrm{OEPx}$. Note, however, that due to the restriction of the OEP to \textit{local} potentials, the obtained energies will be higher than the Hartree--Fock results that allow for non-local potentials.
Furthermore, as it is clear from the previous discussion, the common OEP procedures applied to the exchange energy do not give the correct exchange force density. Instead they will generate a purely longitudinal vector field that is not related to the exchange force density in a direct manner.
We usually loose control over the connection between the energy terms and the corresponding force densities (which leads, among others, to a violation of the virial relation). An important exception is the exchange-only local density approximation, where the connection still holds as can be shown directly.
The same holds for correlation approximation. Any approximate correlation energy can always only lead to a longitudinal vector field via the corresponding (generalized) energy derivative, while an approximation based on forces will usually include a transverse component. This means that there is no strict connection between the energy-based and force-based approximations.
To put it differently, if we want to build approximations in DFT based on Hartree and exchange terms beyond the local-density approximation, we have to decide whether we use the exchange energy \textit{or} the exchange force density. Both strategies will only agree when we use the \textit{exact} exchange and correlation terms together.

In the force-based approach we find an additional exact constraint that holds locally for the transverse component of the exchange vector field and relates it directly to correlation effects. This is through the previous observation that by Eq.~\eqref{eq:PurelyLongitundinal} the $\fHxc$ is purely longitudinal and since the same holds by construction for the Hartree part, we must have a zero transverse contribution in $\fx + \fc$. If we now define $\alphafc$ analogous to $\alphafx$ then this means that at each point in space and for every spin component it must hold that
\begin{equation}\label{eq:alpha-contraint}
    \alphafx(\rr\sigma) + \alphafc(\rr\sigma) = 0.
\end{equation}
To have such an exact constraint that gives direct access to some local correlation effects can be seen as an advantage of the force-based approach over the usual energy-based, global viewpoint.

Finally, let us comment on the homogeneous-density limit.
In \citet{tchenkoue2019force} it was demonstrated how the usual Slater X$\alpha$ \cite{slater1951} and local-density approximation (LDA) \cite{parr-yang-book} formulas for the local-exchange potential can be derived directly from the exchange-force expression Eq.~\eqref{eq-interaction-force-single-slater}. Since $\fx(\rr\sigma)$ is purely longitudinal for a homogeneous density, the exact same derivation can also be started immediately from the local-exchange potential expression of Eq.~\eqref{eq:LocalExchangePotential-1}. A related derivation of the same fact based on the second-order gradient expansion of the exchange-hole density was already given in \citet{wang1990exchange}. This directly connects the most fundamental functional approximations of DFT with the present formalism.

\section{The force-based approach in other DFT variants}
\label{sec:other}

Another advantage of the force-based approach is the inherent compatibility to CDFT and time-dependent DFT (TDDFT). The generalized exchange virial relation Eq.~\eqref{eq:VirialGeneral} highlights the connection of the force-based approach to CDFT. If besides the density $\rho$ we also intend to control the current density $\jj$, then we would need a transverse exchange-correlation vector potential as well, where $\alphafx$ contributes to the exchange vector potential~\cite{tchenkoue2019force}. We even find that $\vfx$ and $\alphafx$ can be chosen to be the local-exchange potential of CDFT and of time-dependent CDFT~\cite{tchenkoue2019force}. This makes $\vfx$ nicely compatible with this variant of DFT.

To make the connection to TDDFT visible, we derive the analogous equation to Eq.~\eqref{eq-Hxc-potential} by subtracting Eqs.~\eqref{eq:force-balance} and \eqref{eq:force-balance-s}, just this time the time-derivative of the currents is not zero.
\begin{equation}\label{eq:TDDFT-KS}
    \partial_t (\jj(\rr\sigma,t) - \jj_s(\rr\sigma,t)) = \rho(\rr\sigma,t) \nabla \vHxc(\rr\sigma,t) + \FHxc(\rr\sigma,t).
\end{equation}
In order to still get rid of the currents, one can apply the divergence and use the continuity equation $\partial_t \rho(\rr\sigma,t) = -\nabla\cdot \jj(\rr\sigma,t) = -\nabla\cdot \jj_s(\rr\sigma,t)$ for both systems that share the density $\rho(\rr\sigma,t)$ at all considered times.
This is how we arrive at~\cite{ruggenthaler2015b}
\begin{align}\label{eq:ExchangeTimeDependent}
    \nabla \cdot \left[\rho(\rr\sigma,t) \nabla v_{\rm Hxc}(\rr\sigma,t) \right] = - \nabla \cdot \FHxc(\rr\sigma,t).
\end{align}
Consequently, the local-exchange potential in TDDFT is now determined from the exchange force density not by solving a Poisson equation but by inverting a Sturm--Liouville equation.
Therefore, the local-exchange potential in TDDFT will be different from $\vfx$, yet the difference can be determined from $\alphafx$~\cite{fuks2018exploring}.
On the other hand, if $\fHxc(\rr\sigma,t)$ would be purely longitudinal then Eq.~\eqref{eq-Hxc-potential} also holds in the time-dependent case and then Eq.~\eqref{eq:ExchangeTimeDependent} is a direct consequence of it by just multiplying with $\rho(\rr\sigma,t)$ and taking the divergence. Conversely, it is only the transverse part of $\fHxc(\rr\sigma,t)$ that makes the difference when we compare \textit{instantaneously} the time-dependent case of Eq.~\eqref{eq:ExchangeTimeDependent} and the static case of Eq.~\eqref{eq-Hxc-potential}. In other words, if we only consider the wavfunctions/forces at a given instant, it is only the non-zero phases/transverse forces that inform us whether we are considering a time-dependent situation. While we do not have access in this \textit{instantaneous} picture to all memory effects~\cite{maitra2002,maitra2016perspective}, we nevertheless see that the transverse forces are important to generate memory over time. In an exchange-only approximation, this role is then taken over by $\alphafx$.

Finally, let us shift attention back to the time-independent setting again. Therein, besides Eq.~\eqref{eq-Hxc-potential}, the exact \emph{ground-state} exchange-correlation potential and force density still also obey Eq.~\eqref{eq:ExchangeTimeDependent}. This gives rise to a different version of the local-exchange potential. Here we will not investigate this alternative force-based formulation further but will compare these different definitions in a forthcoming publication. It however highlights a route to more, possibly useful conditions: higher-order equations of motions bring with them new exact constraints.

\section{Numerical tests}

Finally, we consider the differences between the force-based approach and the energy-based approach in practice, with a focus on the effects of the transverse part of the exchange vector field $\fx$ expressed through the vector potential $\alphafx$.
For this, we solve the KS equation in exchange approximation (FBEx), i.e., we take $\vfx$ from Eq.~\eqref{eq:LocalExchangePotential-1} and assume $\vfc=0$ for the total $\vHxc$ in Eq.~\eqref{eq:HelmholtzHx} in every KS iteration step, and check how this performs in comparison to common exchange approximations.
In this investigation we do not yet employ the transverse part of $\fx$ somehow beneficially. Yet, involving only the longitudinal component of $\fx$ in the calculation is equivalent to considering the full $\fx$ plus the transverse part from $\fc$ since Eq.~\eqref{eq:alpha-contraint} holds as an exact constraint. A force-based approximation focusing purely on exchange effects thus would need to also consider the transverse contribution from the exchange vector field. To summarize, there are two possible viewpoints on this approximation that are equally justified: When considering only the $\vfx$ exchange potential then exchange effects from $\alphafx$ are missing, or alternatively, that this procedure additionally includes correlation effects from $\alphafc$.

\begin{figure}[ht]
  \begin{center}
    \includegraphics[width=\columnwidth]{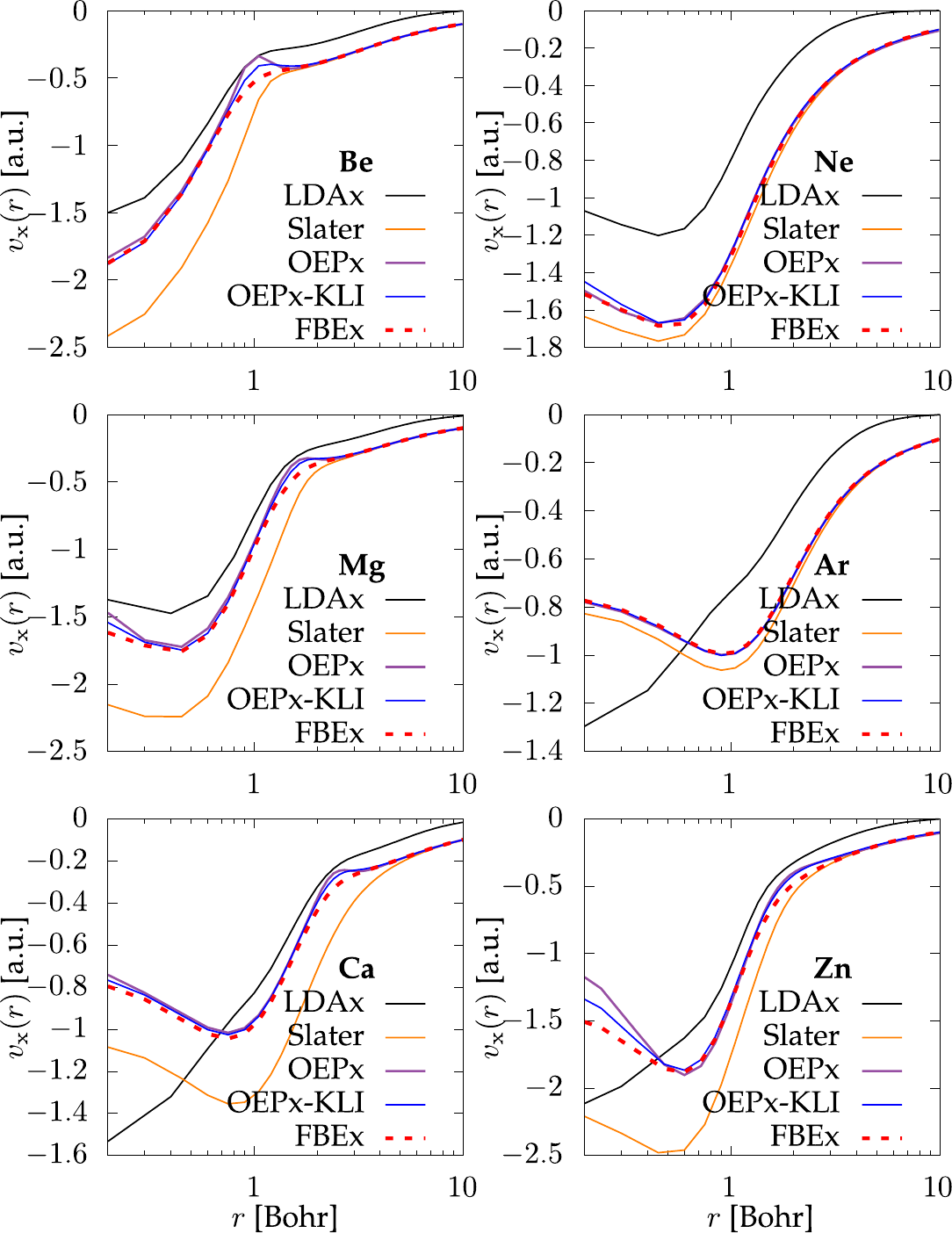}
    \caption{\label{fig:potential} Various local-exchange potentials for different atoms.}
  \end{center}
\end{figure}

We have implemented the force-based local-exchange potential in the real-space code Octopus~\cite{Octopus_paper_2019} and ran simulations for a set of atoms in closed-shell configurations using norm-conserving pseudopotentials~\cite{van2018pseudodojo}, a grid spacing of 0.15 Bohr, and a radius of 10 Bohr for Be and Ne, a radius of 12 Bohr for Mg, Ar, and Zn, and a radius of 14 Bohr for Ca. We found that the FBEx potential performs similar to the much more involved OEP in exchange approximation (OEPx) or its further approximation OEPx-KLI~\cite{PhysRevA.45.101} (see Fig.~\ref{fig:potential}).
While the pure Slater, FBEx and OEPx-KLI potentials all share the same computational scaling as Hartree--Fock, the OEPx method only works as an iterative procedure and is more costly.
Furthermore, we demonstrate that the local-exchange potential adheres to the virial relation of the form of Eq.~\eqref{eq:virial-relation-standard} up to numerical inaccuracies (see Tab.~\ref{tab:table_virial}) because of spherical symmetry, while the OEPx and the OEPx-KLI violate this relation. This numerically confirms that the exchange functional is not functionally differentiable. Further numerical tests and comparisons, also for small molecules, can be found in App.~\ref{app:Numerical}.

\begin{table}[ht]
\begin{ruledtabular}
\begin{tabular}{lccccc}
Atom & Slater & FBEx & OEPx-KLI & OEPx \\
\hline
Li & 245.3 & -0.049 & 0.647 & -1.464 \\
Be & 415.2  & 0.037 & 32.87  & 17.97  \\
Ne & 208.2 & -0.001 & 30.94 & 33.139 \\
Na & 896.1 & 0.080 & -22.08 & -36.4 \\
Mg & 1328.1 & 0.353 & 60.40 & -19.49 \\
Ar & 221.95 & 0.000 & 7.61 & 8.21  \\
Ca & 603.4 & -0.011 & 17.51 & -1.52 \\
Zn & 6225.2 & 0.82 & 26.11 & -83.34 
\end{tabular}
\end{ruledtabular}
\caption{\label{tab:table_virial} Difference $\Delta E_{\mathrm{x}} = E_{\mathrm{x}}^{\mathrm{eig}}-E_{\mathrm{x}}^{\mathrm{virial}}$, in mHa, between the exchange energy computed from the orbitals (or the density) and from the exchange energy obtained from the potential using the virial relation for different local-exchange potentials.}
\end{table}

\begin{figure}[ht]
  \begin{center}
    \includegraphics[width=\columnwidth]{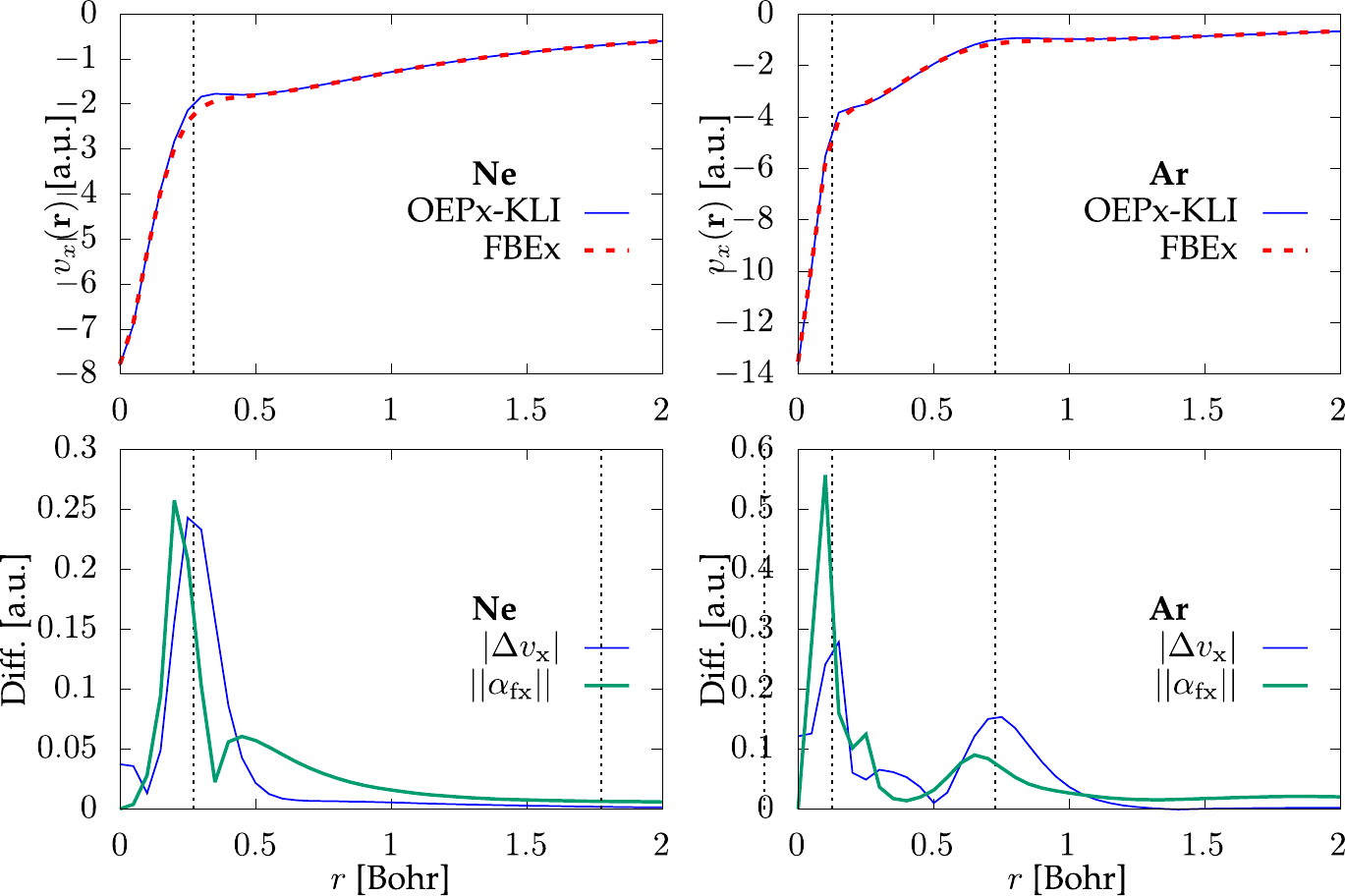}
    \caption{\label{fig:potential_all_electron} Top panels: Similar as Fig.~\ref{fig:potential} for all-electron calculations. Bottom panels: Difference $\Delta \vx$ between $v_{\rm FBEx}$ and $v_{\text{OEPx-KLI}}$, and rescaled norm of $\alphafx$ from the exchange force as in Eq.~\eqref{eq:HelmholtzDecomposition}. The vertical ines indicate the positions of the bumps.}
  \end{center}
\end{figure}

\begin{figure}[ht]
  \begin{center}
    \includegraphics[width=\columnwidth]{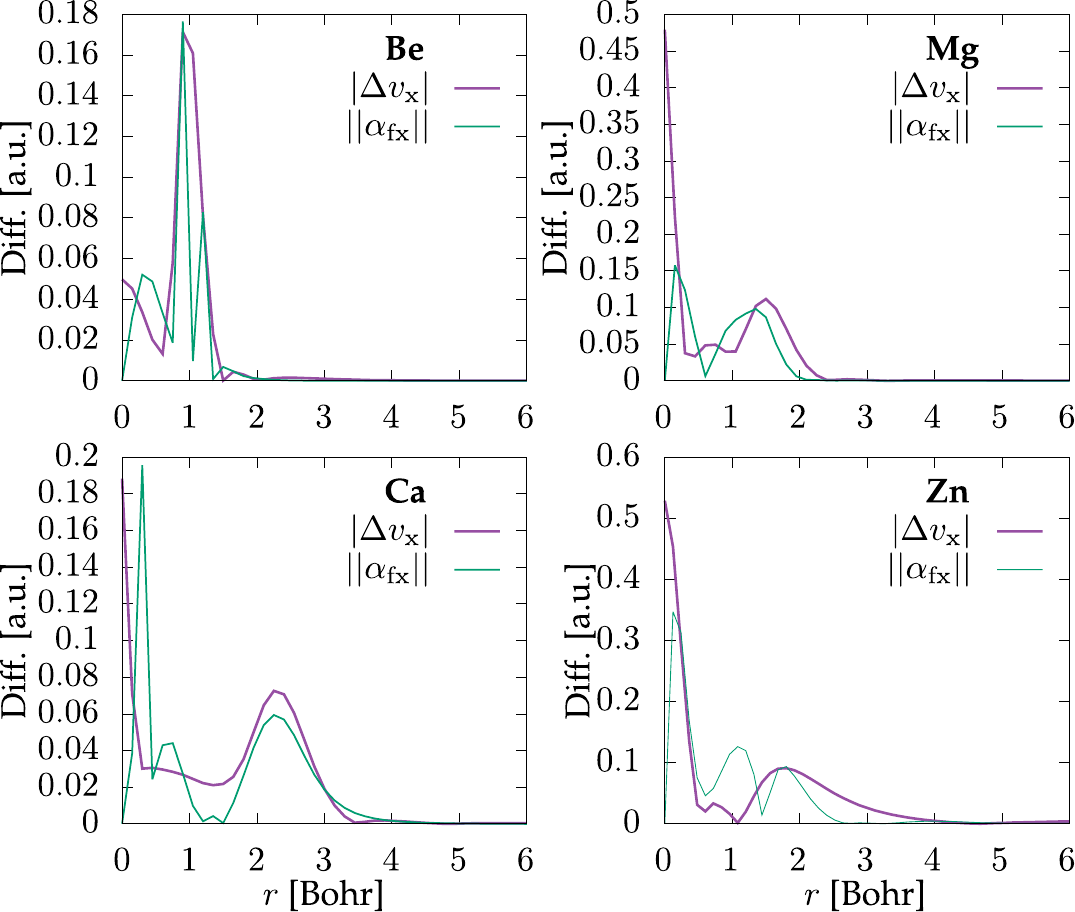}
    \caption{\label{fig:alpha} Difference $\Delta \vx$ between $v_{\rm FBEx}$ and $v_{\rm OEPx}$, and rescaled norm of $\alphafx$ from the exchange force for some of the atoms of Fig.~\ref{fig:potential}.}
  \end{center}
\end{figure}

In fact, Fig.~\ref{fig:potential} shows that the FBEx and the OEPx potentials are almost identical, apart from the small ``bumps'' that are an indication of the shell-structure of the atoms. The Slater potential also does not capture them and a suitable correction for it based on the kinetic-energy density is available \cite{Becke2006}.
Due to the use of pseudopotentials in our simulations displayed in Fig.~\ref{fig:potential}, we see here either one or no bump. To check that the differences between these two potentials are indeed only present at the shells of the atoms, we also performed all-electron calculations for Ne and Ar using a grid spacing of 0.05 Bohr and a radius of 14 Bohr and we obtain that the potentials differ only at the location of the bumps, see the top panels of Fig.~\ref{fig:potential_all_electron}. The lower panels of Fig.~\ref{fig:potential_all_electron} show the difference between OEPx-KLI and FBEx together with $\|\alphafx\|$, i.e., the transverse part of $\fx$.
The same comparison is conducted with OEPx for those atoms where a bump is visible despite using pseudopotentials, see Fig.~\ref{fig:alpha}. In each case we find that the FBEx force has a non-vanishing transverse part only at the position of the bumps, and that the norm of $\alphafx$ follows a pattern similar to the difference between the FBEx and OEPx potentials, clearly showing that there is a connection between the transverse part of the force and the bumps of the OEPx potential.
In fact, we interpret our result in the following way: The bumps appear in the OEPx potential as the procedure tries to impose a longitudinal vector field at places where the exchange vector field actually has a transverse component. Thus, even if we do not employ the transverse part of the forces explicitly, they contain physical information (related to the shell structure of atoms) that can be potentially used for more advanced approximations. For instance, this feature is related to to the correlation forces $\fc$, since $\alphafx$ needs to precisely compensate $\alphafc$. In this manner, we get \textit{local} information about the correlation vector field that could provide quite stringent constraints on future approximations.

A further comparison of the FBEx and OEPx-KLI methods with inversion procedures that yield the full exchange-correlation potential is performed in Appendix~\ref{app:Comp_exact}.

\section{Outlook and Conclusions}

Considering all the different insights obtained by this investigation, we want to highlight two specific results that we deem important for the future of force-based approximations. On the one hand, we have seen that the transverse part of the exchange vector field contains important physical information. It stands to reason that the standard OEP procedure tries to turn these transverse parts into longitudinal contributions of the corresponding OEP potential which are responsible for the appearance of the ``bumps''. An obvious way of including these contributions is to employ an auxiliary system that also contains a vector potential instead of the usual KS system with only a scalar potential. Using the beneficial connection of the force-based approach to CDFT, the corresponding exchange vector field is given via a non-linear partial-differential equation~\cite{tchenkoue2019force}. This paves the way to obtain a semi-locally acting vector potential in the context of electronic ground-state DFT.

On the other hand, for time-dependent DFT, we have seen that the appearance of the transverse vector field implies non-adiabaticity. That is, if we solve the corresponding Sturm--Liouville equation \eqref{eq:ExchangeTimeDependent} instead of the Poisson equation \eqref{eq:HelmholtzHx}, we automatically get a non-adiabatic functional based on force densities. These two aspects make the force-based approach quite promising to find more accurate yet numerically inexpensive approximations within density-functional theories. It is even relatively easy to extend the present approach to other variants of DFT, for instance to forms that include noncollinear magnetism and spin-orbit coupling~\cite{von1972local,PhysRevB.61.15228,PhysRevLett.87.206403,kubler1988density}. Using the corresponding equations of motions for the current density~\cite{tancogne2022effect}, one can apply the same Hartree-exchange and correlation force density decomposition and hence is able derive the corresponding potentials also for this case. Furthermore, in order to address the still unknown correlation force density we highlight that the transverse part of the exchange vector field provides us with \textit{local} constraints on approximate correlation force densities. In the correlation force density, the interaction part $\FF_{W}[\Psi] - \FF_{W}[\Phi]$ can be expressed by the correlation hole, while the kinetic part $\FF_{T}[\Psi] - \FF_{T}[\Phi]$ can be expressed as the difference between the interacting and the non-interacting one-body reduced density matrix close to the diagonal~\cite{fuks2018exploring}. Approximations can then be tested by comparing to the transverse exchange vector field. From this perspective, the success of LDA-based approximations can be explained by the fact that already on the exchange level no transverse forces appear, such that the virial relation is fulfilled, and hence adding purely longitudinal correlations obeys the zero-transverse vector field constraint of Eq.~\eqref{eq:PurelyLongitundinal}. Alternatively, one can start from approximated correlated reduced density matrices and derive the corresponding forces. One can therefore either try to build approximate models based on physical intuition~\cite{giesbertz2013towards}, derive expressions for these terms for specific cases (e.g., the homogeneous limit) from wave-function methods potentially augmented by modern machine-learning techniques~\cite{AI_science}, or devise perturbative expansions on top of the KS Slater determinants.
Even though the force densities are three-dimensional vector fields and thus more involved than energy expressions, the previously successful application of the aforementioned approaches to construct correlation-energy functionals makes it plausible that similar methods are well applicable to the force-based approach to KS-DFT.

In conclusion, we have shown that defining the Hx potential and energy of KS-DFT by forces is not only conceptually beneficial, but also has certain advantages in practice over the common energy-based approach. It is numerically straightforward to construct the corresponding potential from a given force density, the method allows to avoid various problems of the energy-based approach such as determining implicit functional derivatives, and it further provides an explicit form for the local-exchange potential and exchange energy from the exchange force density.
This force-based local (in the sense on how it acts on the wave function) exchange approximation depends \textit{non-locally} on all other points and all occupied orbitals and is numerically as cheap as the Slater potential.
The non-explicit correlation potential is defined uniquely by the correlation force density and in contrast to the energy-based approach, the role of correlations in compensating the transverse part of the exchange vector field is transparent. It is seen that the exchange vector field provides \textit{local} information about the properties of the correlation vector field. We also have a straightforward connection to the current-density variant of DFT and to the time-dependent case. Furthermore, the approach can be seamlessly applied to atomic, molecular and solid-state systems. 
We showed numerically that the well-known bumps of the OEPx potential are connected to the transverse exchange vector field and with this also to the correlation vector field due to the exact constraint that the transverse exchange vector field is exactly compensated by the corresponding correlation effects. We think, following the ideas of John Perdew and others, that such \textit{local} exact constraints are a good starting point to help in devising correlation force density approximations, in DFT and its variants.

\begin{acknowledgements}
This work was supported by the European
Research Council (ERC-2015-AdG694097), by the Cluster of
Excellence ``CUI: Advanced Imaging of Matter'' of the Deutsche
Forschungsgemeinschaft (DFG) -- EXC 2056 -- project ID 390715994,
and the Grupos Consolidados (IT1249-19).
MP, MAC and AL have received funding from the ERC-2021-STG under grant agreement No.~101041487 REGAL. MAC and AL were also supported by  the Research Council of Norway through funding of the CoE Hylleraas Centre for Quantum Molecular Sciences Grant No.~262695 and CCerror Grant No. 287906.
\end{acknowledgements}

%

\onecolumngrid
\appendix

\section{Numerically convenient forms of the local-exchange potential}
\label{app:Convenient}

In order to bring the local-exchange potential into a numerically more convenient form, we perform a partial integration in Eq.~\eqref{eq:LocalExchangePotential-1} and find 
\begin{align}\label{eq:LocalExchangePotential-2}
\vfx(\rr\sigma) = v_{\rm SL}(\rr\sigma) + \int \frac{\nabla' \cdot \int |\rr''-\rr'|^{-1} \nabla' \bar{\rho}_{\rm x}(\rr''|\rr'\sigma)\d\rr''}{4 \pi |\rr-\rr'|}\d\rr',	
\end{align}	
where $v_{\rm SL}(\rr\sigma) = \int |\rr-\rr'|^{-1} \bar{\rho}_{\rm x}(\rr|\rr'\sigma) \d\rr'$ is the Coulomb potential generated by the exchange hole, i.e., the well-known Slater exchange potential. The second term can be computed using \begin{equation}
 \int |\rr''-\rr'|^{-1} \nabla' \bar{\rho}_{\rm x}(\rr''|\rr'\sigma)\d\rr'' = \sum_{ij} \left(\nabla'  \frac{\rho_{ij}^*(\rr'\sigma)}{\rho(\rr'\sigma)} \right) \int |\rr''-\rr'|^{-1} \rho_{ij}(\rr''\sigma) \d\rr'',
\end{equation}
where we defined the co-density $\rho_{ij}(\rr\sigma) = \phi_i^*(\rr\sigma)\phi_j(\rr\sigma)$.

This form has a few advantages. First, we only need to solve one Poisson equation and compute one gradient per pair of indices $i,j$. Therefore, the numerical cost only increases by one gradient per pair of indices $i,j$ compared to the Slater potential. Further, it provides an analytical expression for beyond-Slater approximations and might such serve as the starting point for the development of novel functionals. Finally, from this expression it is also clear that in the single orbital case $ \rho_{ij}^*(\rr'\sigma)/\rho(\rr'\sigma) $ is uniformly equal to 1 and that then the second term vanishes.

There is still a subtle numerical issue when implementing this expression. When evaluating $\rho_{ij}^*(\rr'\sigma)/\rho(\rr'\sigma)$ close to the border of the simulation box we obtain 1, as in the one-electron limit only the highest occupied state contributes to the density. Having zero-boundary conditions at the border of the box leads to a step function irrespective of the size of the simulation box. Consequently, the evaluation of the gradient on the real-space grid by finite differences leads to a non-zero contribution at the surface of the simulation box. This ``surface charge'' leads to a uniform potential which is not physical.
In order to circumvent this issue, we simply used the Leibniz product rule to evaluate
\begin{equation}
 \nabla'  \frac{\rho_{ij}^*(\rr'\sigma)}{\rho(\rr'\sigma)} = \frac{\rho(\rr'\sigma)\nabla'\rho_{ij}^*(\rr'\sigma) - \rho_{ij}^*(\rr'\sigma) \nabla' \rho(\rr'\sigma) }{\rho(\rr'\sigma)^2}\,.
\end{equation}
The numerator is computed first and the two contributions exactly cancel, which leads to the correct long-range numerical value of the potential.

\section{Scaling behavior and virial relation}
\label{app:VirialExchange}

If one uses the coordinate-scaled densities
\begin{equation}
    \rho_\lambda(\rr\sigma) = \lambda^3 \rho((\lambda\rr)\sigma),\quad
    \rho_\lambda^{(1)}(\rr\sigma,\rr'\sigma') = \lambda^3 \rho^{(1)}((\lambda\rr)\sigma,(\lambda\rr')\sigma'),
\end{equation}
one finds $v_{\mathrm{fx},\lambda}(\rr\sigma) = \lambda \vfx((\lambda\rr)\sigma)$, where $v_{\mathrm{fx},\lambda}$ is the expression from Eq.~\eqref{eq:LocalExchangePotential-1} with $\rho \mapsto \rho_\lambda$ and $\rho^{(1)} \mapsto \rho_\lambda^{(1)}$ replaced.
Similarly one finds $\Ex[\rho_\lambda] = \lambda \Ex[\rho]$ directly from Eq.~\eqref{eq:ExchangeEnergy}.
This is the correct scaling behaviour for the exchange energy \cite{parr-yang-book}. 
Together with the assumption of functional differentiability of $\Ex[\rho]$ as a \emph{density} functional and applicability of the usual chain rule, this suffices to derive the virial relation of Eq.~\eqref{eq:virial-relation-standard}. By virtue of the chain rule of functional calculus we have
\begin{equation}\label{eq:Ex-vx-VR-derivation}
\begin{aligned}
    \Ex[\rho] &= \left.\frac{\d \Ex[\rho_\lambda]}{\d \lambda}\right|_{\lambda=1} = \sum_\sigma \! \int \frac{\delta \Ex[\rho]}{\delta\rho(\rr\sigma)} \left.\frac{\d\rho_\lambda(\rr\sigma)}{\d\lambda}\right|_{\lambda=1} \d\rr = \sum_\sigma \! \int \frac{\delta \Ex[\rho]}{\delta\rho(\rr\sigma)} (3\rho(\rr\sigma) + \rr\cdot\nabla\rho(\rr\sigma)) \d\rr \\
    &= - \sum_\sigma \!\int\! \rho(\rr\sigma)\rr\cdot\nabla \frac{\delta \Ex[\rho]}{\delta\rho(\rr\sigma)} \d\rr,
\end{aligned}
\end{equation}
where the last step involves partial integration and the easy identity $\nabla\cdot\rr=3$. It needs to be stressed that this form of the virial relation depends of the assumption of functional differentiability and that $\vx(\rr\sigma) = \delta\Ex[\rho]/\delta\rho(\rr\sigma)$ defines a \emph{different} local-exchange potential than $\vfx(\rr\sigma)$ given by a force-based approach in Eq.~\eqref{eq:LocalExchangePotential-1} or by OEPx.

Next, we prove the virial relation between $\Ex[\Phi]$ and $\Fx[\Phi]$ by direct computation.
Starting with the exchange-energy expression Eq.~\eqref{eq:ExchangeEnergy}, we use the identity $(\rr-\rr')\cdot\nabla|\rr-\rr'|^\alpha = \alpha |\rr-\rr'|^\alpha$ (that is also central for deriving the usual \textit{virial theorem}) with $\alpha=-1$ and the symmetry of the whole expression in $\rr \leftrightarrow \rr'$.
\begin{equation}\label{eq:App-Ex}
\begin{aligned}
    \Ex[\Phi] &= -\frac{1}{2} \sum_\sigma\!\int \frac{| \rho_s^{(1)}(\rr\sigma,\rr'\sigma)|^2}{|\rr-\rr'|} \d\rr \d\rr' = \frac{1}{2} \sum_\sigma\!\int \left((\rr-\rr')\cdot\nabla|\rr-\rr'|^{-1}\right) |\rho_s^{(1)}(\rr\sigma,\rr'\sigma)|^2 \d\rr \d\rr' \\
    &= \frac{1}{2} \sum_\sigma\!\int \left((\rr\cdot\nabla+\rr'\cdot\nabla')|\rr-\rr'|^{-1}\right) |\rho_s^{(1)}(\rr\sigma,\rr'\sigma)|^2 \d\rr \d\rr'= \sum_\sigma\!\int \rr\cdot(\nabla |\rr-\rr'|^{-1}) |\rho_s^{(1)}(\rr\sigma,\rr'\sigma)|^2 \d\rr \d\rr' \\
    & = \sum_\sigma\!\int \rr \cdot \Fx[\Phi](\rr\sigma) \d\rr
\end{aligned}
\end{equation}
Exactly the same relation can be derived for the Hartree energy in an analogous way.
In order to extend the relation towards $\vfx$, in Eq.~\eqref{eq:LocalExchangePotential-1} we first switch $\nabla'$ over to the term $1/(4 \pi |\rr-\rr'|)$ by partial integration and then switch $\nabla' \to -\nabla$ by symmetry.
\begin{equation}
    \vfx(\rr\sigma) = \nabla \cdot \int \frac{(\nabla' |\rr''-\rr'|^{-1}) }{4 \pi |\rr-\rr'|} \frac{|\rho_s^{(1)}(\rr'\sigma,\rr''\sigma)|^2}{\rho(\rr'\sigma)}\d\rr'\d\rr''
\end{equation}
Now putting this into the right hand side of a virial relation of the type of Eq.~\eqref{eq:Ex-vx-VR-derivation} we get
\begin{equation}
\begin{aligned}
    - \sum_\sigma \!\int\! \rho(\rr\sigma) \rr \cdot \nabla \vfx(\rr\sigma) \d\rr 
    =& -\sum_\sigma \!\int\! \rho(\rr\sigma) \rr \cdot \nabla \left( \nabla \cdot \frac{1}{4 \pi |\rr-\rr'|} (\nabla' |\rr''-\rr'|^{-1}) \right) 
    \frac{|\rho_s^{(1)}(\rr'\sigma,\rr''\sigma)|^2}{\rho(\rr'\sigma)} \d\rr \d\rr'\d\rr'' \\[0.6em]
    =& -\sum_\sigma \!\int\! \rho(\rr\sigma) \rr \cdot \underbrace{\left( \Delta \frac{1}{4 \pi |\rr-\rr'|} \right)}_{-\delta(\rr-\rr')} (\nabla' |\rr''-\rr'|^{-1}) \frac{|\rho_s^{(1)}(\rr'\sigma,\rr''\sigma)|^2}{\rho(\rr'\sigma)} \d\rr \d\rr'\d\rr'' \\[-0.6em]
    &-\frac{1}{4\pi} \sum_\sigma \!\int\! \rho(\rr\sigma) \rr \cdot \left( \nabla \times \left( (\nabla |\rr-\rr'|^{-1} ) \times (\nabla' |\rr''-\rr'|^{-1}) \right)\right) \frac{|\rho_s^{(1)}(\rr'\sigma,\rr''\sigma)|^2}{\rho(\rr'\sigma)} \d\rr \d\rr'\d\rr'',
\end{aligned}
\end{equation}
where the vector calculus identities $\nabla (\nabla \cdot \AA) = \Delta \AA + \nabla \times (\nabla \times \AA)$ and $\nabla \times f(\rr) \mathbf C = (\nabla f(\rr))\times  \mathbf C$ were used.
Now the first part gives exactly $\Ex$ according to Eq.~\eqref{eq:App-Ex} while the second line appears as an additional term in a virial relation between $\Ex$ and $\vfx$. But since it appears as the curl of a vector expression it cannot be equal to the gradient of a scalar potential, so the difference comes from the transverse part of $\fx$ while $\vfx$ corresponds only to the longitudinal part of $\fx$. The nice thing is that this gives an explicit form for the transverse part of $\fx$, while the longitudinal part is already given by $-\nabla \vfx$. We thus find the following Helmholtz decomposition,
\begin{align}
    &\fx(\rr\sigma) = \frac{\Fx[\Phi](\rr\sigma)}{\rho(\rr)}= -\nabla \vfx(\rr\sigma) + \nabla \times \alphafx(\rr\sigma), \\
    &\vfx(\rr\sigma) = \frac{1}{4\pi} \int (\nabla' |\rr-\rr'|^{-1}) \cdot (\nabla' |\rr''-\rr'|^{-1}) \bar{\rho}_{\rm x}(\rr''|\rr'\sigma)\d\rr'\d\rr'', \\
    &\alphafx(\rr\sigma) = \frac{1}{4\pi} \int (\nabla' |\rr-\rr'|^{-1}) \times (\nabla' |\rr''-\rr'|^{-1}) \bar{\rho}_{\rm x}(\rr''|\rr'\sigma)\d\rr'\d\rr'',
\end{align}
and the extended virial relation
\begin{equation}\label{eq:virial-with-vector-pot}
    \Ex[\Phi] = \sum_\sigma \!\int\! \rr \cdot \Fx[\Phi](\rr\sigma) \d\rr = -\sum_\sigma \!\int\! \rho(\rr\sigma)\rr \cdot \nabla \vfx(\rr\sigma) \d\rr + \sum_\sigma \!\int\! \rho(\rr\sigma)\rr \cdot (\nabla \times \alphafx(\rr\sigma)) \d\rr.
\end{equation}
If in certain situations it holds that the second term above is zero then the virial relation between $\Ex[\Phi]$ and $\vfx$ holds in the form of Eq.~\eqref{eq:Ex-vx-VR-derivation}.
We show that for spherically-symmetric densities $\rho(\rr\sigma)=R_\sigma(|\rr|)$ this is indeed the case. For this we take the last integral of Eq.~$\eqref{eq:virial-with-vector-pot}$ and perform integration by parts with the curl and vanishing boundary terms to get
\begin{equation}
    \sum_\sigma \!\int\! \rho(\rr\sigma)\rr \cdot (\nabla \times \alphafx(\rr\sigma)) \d\rr = -\sum_\sigma \!\int\! (\nabla \times \rho(\rr\sigma)\rr) \cdot \alphafx(\rr\sigma) \d\rr = -\sum_\sigma \!\int\! (\rho(\rr\sigma)(\nabla \times \rr) + (\nabla \rho(\rr\sigma)) \times \rr) \cdot \alphafx(\rr\sigma) \d\rr.
\end{equation}
But now $\nabla\times\rr=0$ and $(\nabla \rho(\rr\sigma))\times\rr = (\rr\times\rr) R_\sigma'(|\rr|)/|\rr|=0$, so the above expression evaluates as zero.

\section{Numerical results for the FBEx functional}
\label{app:Numerical}

Here, we show further numerical comparisons of the force-based local-exchange potential to well-established exchange potentials in DFT. Firstly, we investigate how the force-based local-exchange potential compares to the Hartree--Fock energies. Since the force-based local-exchange potential is not derived directly from the exchange-energy expression of Eq.~\eqref{eq:ExchangeEnergy}, it is not designed to approximate the non-local Hartree--Fock exchange-energy expression. Still, the resulting energies of the force-based local-exchange potential determined from Eq.~\eqref{eq:ExchangeEnergy} together with the respective orbitals (see Tab.~\ref{tab:table_HF}) are in good agreement with the Hartree--Fock exchange energies. Note that due to the nonlinear core correction from the pseudopotential and the larger number of valence electrons, the results for Zn show a larger discrepancy with Hartree--Fock.

\begin{table}[ht]
\begin{ruledtabular}
\begin{tabular}{lcccc}
Atom & Slater & FBEx & OEPx-KLI & OEPx \\
\hline
Li & -29.34 & 1.496 & 1.234 & 0.787 \\
Be & -39.33 &-2.255  & 0.040 & 0.909 \\
Ne & -27.51 & -7.411 & -1.981 & 2.505 \\
Na & -98.07 & 2.112 & 2.756 & 4.400 \\
Mg & -118.2 & -2.106 & -0.467 & 4.099\\
Ar & -22.34 & 2.417 & 0.856 & 0.081 \\
Ca & -91.12 & 0.113 & 1.903 & 1.508 \\
Zn & -365.7 & -81.22 & 56.42 & 9.788\\\hline
MARE(\%) & 1.49 & 0.116 & 0.077 & 0.035 
\end{tabular}
\end{ruledtabular}
\caption{\label{tab:table_HF} Deviation from the Hartree--Fock exchange energy, in mHa, for different exchange functionals. We also report the mean absolute relative error (MARE) for each functional.}
\end{table}

In Tab.~\ref{tab:table3}, we further report the eigenvalue of the highest occupied orbital for different exchange functionals. While OEPx-KLI and OEPx are yielding similar ionization energies as Hartree--Fock, within a meV precision, the force-based local-exchange potential leads to only slightly different results. The Slater potential shows a stronger deviation from the Hartree--Fock values. 

\begin{table}[ht]
\begin{ruledtabular}
\begin{tabular}{lccccc}
Atom & Slater & FBEx & OEPx-KLI & OEPx & HF \\
\hline
Li & 0.101 & 0.086 & 0.082 & 0.082 & 0.082 \\
Be & 0.325 & 0.311 & 0.307 & 0.307 & 0.307 \\
Ne & 0.900 & 0.859 & 0.843 & 0.845 & 0.844\\
Na & 0.118 & 0.083 & 0.074 & 0.074 & 0.074 \\
Mg & 0.285 & 0.260 & 0.253 & 0.253 & 0.253 \\
Ar & 0.622 & 0.585 & 0.590 & 0.590  & 0.590 \\
Ca & 0.224 & 0.201 & 0.195 & 0.195 & 0.195\\
Zn & 0.368 & 0.332 & 0.300 & 0.300 & 0.300
\end{tabular}
\end{ruledtabular}
\caption{\label{tab:table3} Eigenvalues $-\epsilon_N$, in Ha, of the highest occupied orbitals for different functionals.}
\end{table}

Tab.~\ref{tab:table_virialII} lists the difference in exchange energy computed from the orbitals and the virial relation of Eq.~\eqref{eq:virial-relation-standard} for small molecules. This shows that for non-spherically-symmetric systems, this virial relation is not respected by the force-based local-exchange potential either. For N$_2$, we employed a N-N distance of $1.09769\Angstrom$. For CO$_2$, we considered a C-O bond length of $1.16\Angstrom$. For CH$_4$, we considered a C-H bond length of $1.087\Angstrom$. In all cases, we employed a grid spacing of 0.15 Bohr and a simulation box made of atom-centered spheres of radii 12 Bohr.

\begin{table}[ht]
\begin{ruledtabular}
\begin{tabular}{lccccc}
Molecule & Slater & FBEx & OEPx-KLI & OEPx \\
\hline
N$_2$ & 93.34 & -134.7 & -276.4 & -235.4 \\
CO$_2$ & 13.64 & -520.9 & -1157 & -660.1 \\
CH$_4$ & 72.36 & -12.517 & -38.66 & -19.15 \\
\end{tabular}
\end{ruledtabular}
\caption{\label{tab:table_virialII} Same as Tab.~\ref{tab:table_virial}, but for small molecules.}
\end{table}

The corresponding ionization energies for these molecules are given in Tab.~\ref{tab:mol_ionization}. Similar to the atomic case, we find that the force-based local-exchange potential performs much better than the Slater potential and yields ionization energies close to the ones obtained from Hartree--Fock or OEPx.

\begin{table}[ht]
\begin{ruledtabular}
\begin{tabular}{lccccc}
Molecule & Slater & FBEx & OEPx-KLI & OEPx & HF \\
\hline
N$_2$  & 0.635 & 0.607  & 0.629 & 0.630 & 0.617 \\
CO$_2$ & 0.619 & 0.586 & 0.545 & 0.544 & 0.546 \\
CH$_4$ & 0.566 & 0.540 & 0.543 & 0.545 & 0.546\\
\end{tabular}
\end{ruledtabular}
\caption{\label{tab:mol_ionization} Same as Tab.~\ref{tab:table3}, but for small molecules.}
\end{table}

\section{Comparison with the exact exchange-correlation potential}
\label{app:Comp_exact}

\begin{figure}[ht]
  \begin{center}
    \includegraphics[width=0.9\columnwidth]{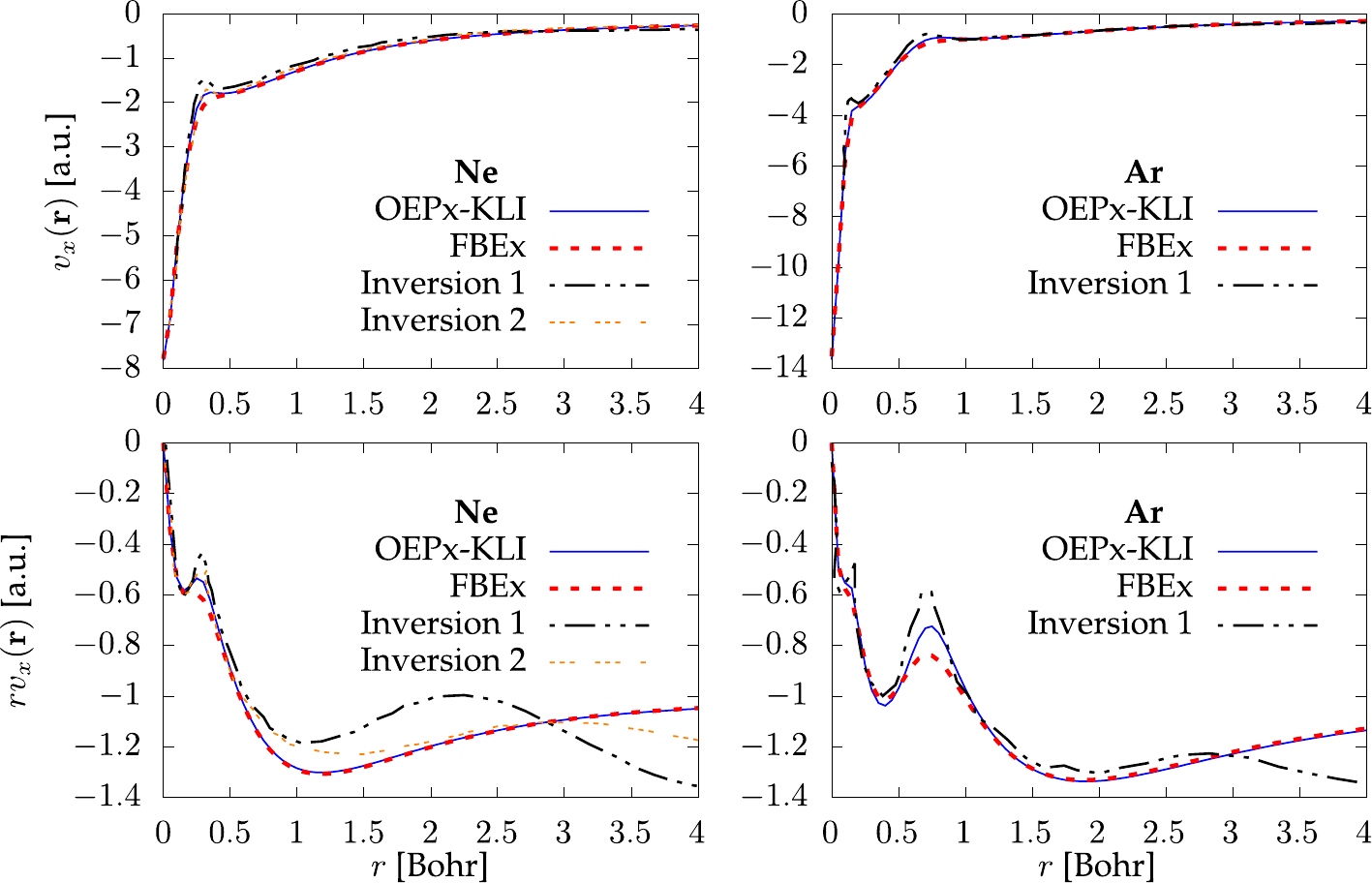}
    \caption{\label{fig:comp_exact} Top panels: Comparison between $v_{\rm FBEx}$, $v_{\text{OEPx-KLI}}$, and exact exchange-correlation potentials $v_{\rm xc}$, for Ne (left panels) and Ar (right panels). The bottom panels shows the potentials multiplied by $r$. ``Inversion 1'' and ``Inversion 2'' refer to the results of the Kohn--Sham inversion procedure from Ref.~\onlinecite{PhysRevA.67.012505} and Ref.~\onlinecite{PhysRevA.49.2421} respectively.}
  \end{center}
\end{figure}

The fact that we only include the longitudinal part of the exchange force density can be viewed as implicitly including a correlation force density that imposes Eq.~\eqref{eq:alpha-contraint}. It is therefore interesting to compare not only to OEPx results, but also to the exact exchange-correlation potential. For the atoms considered in the present work, this was done by the mean of the Kohn--Sham inversion procedure, for instance based on Green's function densities,~\cite{PhysRevA.67.012505} or from CI densities.~\cite{PhysRevA.49.2421}
The comparison are shown in Fig.~\ref{fig:comp_exact}. From these results, it is clear that the implicitly included correlation part does not seem to agree the agreement with the exact potential, as the bumps representing the atomic shells are still a dominant feature in this potential.

\end{document}